# Quantitative Analysis of DoS Attacks & Client Puzzles in IoT Systems

Luca Arnaboldi and Charles Morisset

School of Computing, Newcastle University, Newcastle upon Tyne, UK
{l.arnaboldi,charles.morisset}@ncl.ac.uk

**Abstract.** Denial of Service (DoS) attacks constitute a major security threat to today's Internet. This challenge is especially pertinent to the Internet of Things (IoT) as devices have less computing power, memory and security mechanisms to mitigate DoS attacks. This paper presents a model that mimics the unique characteristics of a network of IoT devices, including components of the system implementing 'Crypto Puzzles' - a DoS mitigation technique. We created an imitation of a DoS attack on the system, and conducted a quantitative analysis to simulate the impact such an attack may potentially exert upon the system, assessing the trade off between security and throughput in the IoT system. We model this through stochastic model checking in PRISM and provide evidence that supports this as a valuable method to compare the efficiency of different implementations of IoT systems, exemplified by a case study.

## 1 Introduction

A DoS attack targets the availability of a device or network [9], with the intent of disrupting system usability. The most common method is referred to as Flooding DoS [10], and may be used as an attempt to deplete the devices' resources including memory, bandwidth and/or battery. A DoS attack against an IoT network has the potential to be significantly more detrimental than one against a standard network. This increased vulnerability is due in part to the low computational power and battery power characteristic of IoT devices [13].

The extant literature has delineated several potential approaches that may be effective in the mitigation of a DoS attack [14]. This paper focuses on one such method, known as "Client Crypto Puzzles"[2], one of the most common mitigation techniques. We evaluate the probability of the system (or subsystem) being denied within a specific time frame in an IoT network. Using our proposed model we are able to assess properties such as: i) At what point is the mitigation technique doing more harm than good? ii) How does denial of a single device impact functionality of the entire system? iii) Does it create a snowball effect?

Client puzzles take many forms, but the general purpose is to force the sender to perform a computationally intensive task prior to authentication, consequently reducing their ability to spam messages [2]. Client puzzles have been adapted in IoT systems [7] and have been shown to successfully decrease effectiveness of a DoS attack. It is however of high computational intensiveness. In the specific context of IoT, an additional consideration is that a client puzzle (especially one of high complexity) can place strain on the battery, causing high delays in throughput whilst a client is occupied with solving the client puzzles. If the puzzle drains battery at a rate equivalent (or more) than a flooding attack, the increased security may actually harm the system. In this paper we model the tradeoff between security and throughput in addition to the impact the increase in computation has on a device's battery life span. It is also the particular case where in the scope of the IoT one device being denied will not harm the system as one device may be performing an inconsequential or very small task. We model this through the connectivity of devices. We can observe the potential snowball effect of a system as the denial of one particular device will increase the probability that other devices are also going down. Through the model we can observe the scenario where DoS mitigation, throughput and decrease in battery are at optimal balance to obtain the best possible result in all three cases. From this, we may model the ideal setup given specific number of devices, connectivity of the devices and DoS strain. In particular we demonstrate that in some cases mitigation techniques can actuallt increase the likelihood of a DoS, due to drainage of battery.

The aim of this research is to quantify the potential impact of a DoS attack on a multi protocol network within the IoT and to gauge how a potential mitigation method affects performance. The key contributions include:

1. A model of two types of IoT device networks, one with DoS mitigation in place and one without.
2. Verification and simulation of these networks to investigate trade-off between security and throughput under a DoS attack.

The remainder of this paper is divided into the following sections; section 2- discusses other work concerned with quantifying impact of DoS and the IoT; section 3 - describes our model; section 4 - details our analysis technique; section 5 - outlines the setup for our experiments; section 6 - highlights findings and results; and Section 7 concludes the paper, summarising our contribution and ongoing work.

## 2 Related Work

Since their advent, systems security properties have been modeled and verified using a variety of tools including probabilistic model checking. Analysis of DoS mitigation techniques has been widely covered, Tritilanunt et al. [15] used colored petri net to verify the effectiveness of HiP client puzzles for DoS mitigation. They mainly used simulation under different scenarios of possible DoS attacks and proposed techniques to predict DoS attacks in advance.

Similarly Basagiannis et al. [5] also looked at HIP, making use of verification techniques. They used probabilistic model checker PRISM, introducing a probabilistic attacker model to analyze the effectiveness of HIP and created different attack paths to break the DoS mitigation technique. Their work focuses on a single complete exchange between an initiator and respondent, creating a Dolev-Yao-like attacker.

Several papers address modeling IoT, adopting various different approaches. Authors of [3, 12] have worked on modeling a specific IoT protocol on the transport layer, looking at MQTT and CoAP respectively. Fruth [6] examines various properties of a Wireless Network protocol including connectivity and energy power through PRISM, and evaluates it on a Wireless Sensor Network System. The author evaluates the battery drainage of certain randomized protocols.

Throughput vs security is a common research question in analysis network. Abdelhakim et al. [1] present work on this particular topic in the context of wireless sensor networks. Their paper introduces a concept of security routing, optimized with throughput to present optimum routing.

An abundance of research exists on effectiveness of client puzzles and throughput vs security. However when we have devices actually going down due to battery strain we observe a phenomenon of snowball effect, and as a consequence the manner in which we implement the network needs to change. To the best of our knowledge we are the first to combine concepts of IoT Systems, restricted resources of IoT devices and DoS attacks in a probabilistic model checking environment.

## 3   Model for IoT Devices

To implement the model we made use of PRISM Model Checker [8] and Continuous Time Markov Chains (CTMC) to abstract systems of IoT devices. The PRISM tool also allows use of statistical model checking, a technique which is particularly effective since modeling multiple devices exacerbate the state space immensely. We ran a series of experiments using different parameters, including DoS attacks against different device setups and different parts of the network. The choice to use CTMC was due to their stochastic properties and the fact that events occur spontaneously, resulting in a wider range of scenarios. A DoS attack is unstoppable given enough time and resource sand as such we deemed time to be a key factor in our analysis. To calculate the likelihood of specific scenarios taking place we use the PRISM verification tools and Probabilistic Computational Tree (PCTL). PCTL is a probabilistic extension to Computational Tree Logic[4] and provides means to evaluate behavior of the system. The model abstracts an IoT network under DoS strain, and it is implemented as a system of *devices*.

A device is a sensor connected to the internet with its own power supply. A key aspect of the IoT is that different devices might have different battery lives and different security features (in this paper, we focus on whether a device is implementing a DoS mitigation technique). Hence, we consider the following device properties: a battery life, a message queue and connectivity (what other devices it can connect to). Battery life is a measure that is drained whenever a computationally intensive operation takes place such as sending a message and computing a client puzzle. A device can only hold a limited number of messages, and after the queue reaches capacity it cannot receive more until it has replied with acknowledgments. If a recipient device queue is full the initiator waits until it either frees up or timeouts and then resends. To model connectivity each device has a set of other devices it can communicate with and receive messages from. To implement the concept of processing time we implemented arbitrary delays when processing client puzzles.

We introduce the concept of "gentlemen devices" and "rude devices". A gentleman device will utilize "politeness", i.e. they will send a message and wait for an acknowledgment (or timeout if acknowledgment takes too long) from the recipient and conclude his current discussion with the device before initiating another message exchange with the same device. It can however simultaneously hold exchanges with other devices. Rude devices on the other hand may continue sending messages to devices within their connectivity before the full communication is over, replicating a flooding attack. The effects of these rude devices flooding spreads as even gentleman devices connected to the flooded device will not be able to commence an exchange if its message queue is full.

The attacker or rude device can have different rates of attack, mimicking different strengths of a DoS attack. We assume it sits outside the network and is not part of the connectivity so it can target any node, but gentleman devices cannot perform an exchange with it. To further simulate the attack strength, at each attack a proportional amount of battery is drained depending on the attack's intensity (rate). We use stochastic probabilities to target random parts of the system as we assume an attacker has no knowledge of the system setup. Due to the connectivity element an attack on one part of the system may exert a higher impact than an attack on another part e.g. If devices B and C only communicate with A, the denial of A stalls the whole system whilst the denial of B still allows the system to function.

Formally, a rude device R is a tuple $R = (S, s_{init}, A, R, L)$, where $S = \{idle, active\}$ is a set of states, $s_{init} = idle$ is the initial state, $A = \{msg, ack\}$ is a set of actions, $R$ is a matrix containing the rates at which any of the actions are performed, e.g., $\begin{pmatrix} 0 & 1 \\ 1 & 0 \end{pmatrix}$ shows there is a rate of 0 to go from *idle* to *idle*, a rate of 1 to go from *idle* to *active*, a rate of 1 to go from *active* to *idle* and a rate of 0 to go from *active* to *active*, finally $L$ is an atomic proposition defined as guard→action where the guard must be true in order for the action to take place and take the attacker into the next state.

The behavior of a rude device is defined as follows: if the device is in state *idle*, there is a probability to move to the state *active*; if the device is in state *active*, the attacker chooses a random node in the network and starts flooding it. If that particular part of the network has mitigation technique in place at each message it has to solve the crypto hash before sending again. The attack continues until the devices battery has been drained. The attacker then goes back to idle. The guard is used to make sure the attacker behaves within the scope of the model. The first atomic proposition assures that the attacker does not target multiple parts of the network simultaneously and then switches to active and the second guard follows the steps to fill the message queue and drain the battery.

The other nodes in the network or gentleman devices may either have DoS mitigation techniques or not. A gentleman device $G$ is formally defined as $G = (S, s_{init}, A, R, L)$, where $S = \{idle, sending, receiving\}$ is a set states, $s_{init} = idle$ is the initial state, $A = \{msg, ack, challenge\}$ is a set of actions, $R$ is a matrix containing the rates at which any of the actions are performed, e.g., $\begin{pmatrix} 0 & 1 & 1 \\ 1 & 0 & 0 \\ 1 & 0 & 0 \end{pmatrix}$ shows there is a rate of 0 to go from *idle* to *idle*, a rate of 1 to go from *idle* to *sending* as well as a rate of 1 to go to *receiving*, a rate of 1 to go from *receiving* to *idle* and a rate of 1 to go from *sending* to *idle*, and $L$ is an atomic proposition defined as guard→action, there are guards to enable the correct behavior of message exchange (e.g. *idle*, A *to* B, B *to* (ACK) A, *idle*).

The behavior of a gentleman device without mitigation technique is as follows: when *idle*, G is active and has a chance to begin a communication between any of his connected devices. From idle it can transition to any of two other states *sending* and *receiving*. If *sending*, G sends a message to a connected device and the battery is drained, it then initiates a timer, if the acknowledgment is not received before the timer runs out the device goes back to *idle* however the reset drains the battery, if it is received it finishes the exchange and also resets to *idle*. If *receiving*, the message is added to the queue and the initiator is noted as to direct the ACKs to the right initiator (multiple messages may be received at the same time). The acknowledgment is then sent and the device is reset to *idle* and the queue is decreased. To implement a device with mitigation techniques, we add the following properties to a gentleman device; if in state *receiving* before it can send the acknowledgment to a initiator with client puzzles there is a time delay to portray the time it would take to solve a puzzle. The time delay increases when the size of K increases, as it mimics how a harder puzzle is more difficult to resolve. We refer the reader to [2] for some examples of client puzzles.

## 4 Verification

To test our model we ran a variety of experiments with different systems and security setups. One of the key aspects of our experiment was how the impact of the DoS attacks scaled with different attack rates, different setups and how it effected the throughput and security to investigate the viability of these mitigation techniques in the context of the IoT. The ideal scenario is when the probability of being denied within time T is low and the throughput after T seconds is high. Observing these circumstances in finite systems where there is a set number of devices, we looked at all possible setups the system could take in terms of how many devices are protected and by what level of client puzzles, and then tested them under different DoS strains. Using the result we can tell which setup is the best suited to a particular rate of DoS and which setup will maximize throughput and security. We theorized that given the circumstances of the IoT and the relatively high processing times at certain levels of puzzle difficulty it would be the case that the lowering in chance of denial would not be as significant as the corresponding lowering in throughput caused by the processing delays, this is analyzed in the case study provided in Section 6.1.

We made use of statistical model checking when examining the larger models through PRISM's simulation engine. This approach is particularly useful on very large models when normal model checking is infeasible. Essentially, this is achieved by sampling: generating a large number of random paths through the model, evaluating the result of the given properties on each run, and using this information to generate an approximately correct result within a specific Confidence Interval (CI). Let X denote the true result of the query $P =?[...]$ and Y the approximation generated. The confidence interval is $[Y - w, Y + w]$, i.e. w gives the half-width of the interval. The confidence level, which is usually stated as a percentage, is $100(1 - Confidence)\%$. This means that the actual value X should fall into the confidence interval $[Y - w, Y + w]100(1 - Confidence)\%$ of the time [11]. We tested for the following properties.

**Throughput:** the number of messages processed over a given time interval (cumulative messages sent/current time). By definition if a message takes longer to send the throughput will decrease, hence adding computationally intensive tasks that delay the transmittance of messages is going to decrease the systems throughput. However if they delay the likelihood of devices being taken down by an attacker the theory is that in the long run it will actually increase the throughput under DoS attacks. To calculate the throughput of the IoT system we make use of PCTL formula $R\{Msg\_sent\} =?[C <= T]$ or what is the cumulative total messages sent by the system in time T and then divide the answer by the value T. The value $Msg\_sent$ is a reward structure that assigns one reward every time a successful message exchange is completed.

**Likelihood of System being denied:** we defined the denial of a device, when either its battery has been completely drained or its connected devices have been drained and it therefore cannot transfer from the *idle* state. The whole system is down when all devices have been *denied*. This is once again monitored through PCTL over a restricted time, running different variables one can optimize the number of protected devices as well a what strength of protection to optimize the implementation least likely to be denied.

**Snowball effect due to denial:** of further interest is the ability to recognize the *critical* sectors of a system. We defined a *critical* sector by examining the impact it's denial has on the rest of the system. Highly critical sectors are also the parts of the system which require more securing. We achieve this by measuring the *snowball effect* or rather after the denial of device A what is the new probability of the rest of the system being down. Theoretically highly *critical* devices will increase the probability substantially whilst non *critical* devices will make a small difference.

To observe the best case scenario of a particular setup, we define a setup as a particular spread and strength of mitigation techniques on a given configuration of devices. We observe the balance between security and throughput, the higher the ratio the better the setup for that particular DoS strength. To observe and test the initial hypothesis in Section 6.1 we create a case study on a specific scenario and evaluate the results for each possible implementation given specific assumptions.

## 5 Experiment Setup

We have used the model described above in PRISM Model Checker as well as PCTL to perform both quantitative verification and simulations, various properties where checked and the model was tested under different scenarios. For verification the full state space is explored whilst for simulation we use the following setup. **Number of Samples:** 100,000, **CI:** 0.001 and **Maximum Path Length:** 1,000,000. For the purpose of the model we were interested in evaluating two key properties: throughput and probability of denial.

We can also examine which part of the subnetwork has been denied first and specifically the time required for the denial to take place. The setup for each experiment was: **Client Puzzle Difficulties (K size):** 5 to 20s, **Time of system (Seconds):** 20 to 200s, **Devices in Network (All protected):** 9 and **Rude Devices:** 1. This differentiated from the case study where all the variables were tested in all possible setups (with some specifications explained in section 6.1) to find the optimal setup.

Results from the initial setup highlight the key factors of our model i) the way client puzzles drain battery ii) the effectiveness of mitigation techniques to avoid denial of service and iii) demonstrating that in finite time it can sometimes be useful to have a less intensive client puzzle as they can disrupt more than help.

## 6 Results

We demonstrated the potential strain that client puzzles place on a system and as can be seen in **Fig. 1** at lower times (i.e. before any part of the systems are under DoS or haven't gone down yet), the higher client puzzles create such a strain that they increase the likelihood of going down rather than diminish it.

It can be observed that the harder the puzzle, the lower the throughput, whilst the security is increased. Furthermore, due to the extra drainage in power of the more difficult puzzle, a smaller value of K performs better in earlier times (At time 80s we observe that K=20 has a higher denial probability than K=15 this is due to the extra processing strain and battery drainage of the harder puzzle). As can be observed towards time 200s the throughputs come to a stall as different parts of the system are denied. It is to be noted that as the lower puzzle difficulties are declining in throughput, K=20 is on the rise, as the increase in security allows for some nodes to still output messages.

### 6.1 Case Study

We apply our model to a specific scenario to demonstrate its application. We assume a potential system engineer views their IoT network as under constant

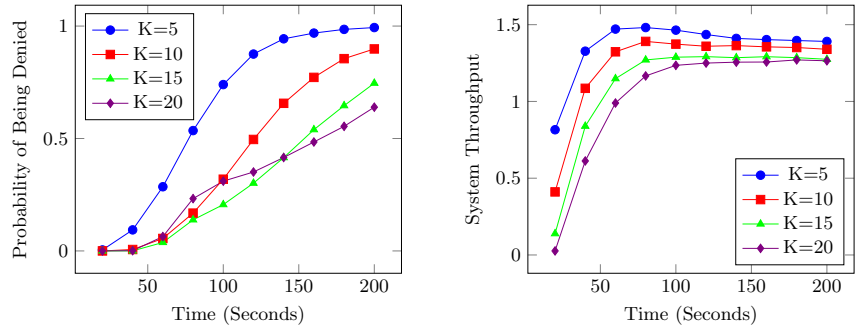

**Fig. 1.** The graphs represents a system being targeted by a DoS attack, the one on the left displays the probability of a DoS attack being successful over time (20s to 200s). The graph on the right represents the throughput of the system over time (20s to 200s).

DoS attack. They wishes to know what would be the best way to optimize the balance between throughput of their system and the likelihood of being denied when implementing security mitigation for these DoS attacks. Furthermore, their IoT system is collecting very critical data in a short time frame so it is important for the system to work at its best for the initial 100s. The setup is the following; A is connected to B, B is connected to A, C is connected to A, B and C, and D is connected to C. We also assume a rude device E, which is connected to all devices.

We ran every single scenario of setups (A protected, AB protected, B protected...) and for each scenario tried every single possible value of K. We set the value of Battery to be 50, a maximum message queue of 5, Time at 100s, a single rude Device (E) with rate of attack the same as the rate of any gentleman device (ABCD) and the values of K from 5 to 15 tested on every protected device, the results can be observed in **Table 1**. If we use our formula of throughput/probability of denial, with the highest value being the most optimal result. We see that the scenario AD provided the best ratio at every different value of K, with scenarios including C being connected (the one with the most connectivity) not performing as well.

The results support our initial hypothesis that mitigation techniques actually increase the likelihood of denial through battery drainage alone. This is evidenced by the scenarios which had C protected that proved to be the most inefficient (throughput dropped significantly). The rude devices have an equal chance of attacking as the gentleman devices. As a consequence, since every single connection to a protected device will drain the battery and since there is equal chance a rude device will start an attack, the strain on the battery caused by normal message sending (and computing of puzzles) is more detrimental. However if we take ACD which theoretically protects the network on all levels and D which is not very connected and test both these two under a longer period of time (200s) we can see the results are altered, as the attackers will continue until a device is denied. On the other hand, if it takes longer to take down a device, the overall system lasts longer and therefore the throughput is higher with the protected devices. We also show that identifying single *critical* devices and protecting them rather than the whole system can be a valuable technique. These results exemplify the potential application and benefit of these techniques.

**Table 1.** Every Scenarios For each setup of Protected Devices (PD) and Unprotected Devices (UD), each setup has a single rude device E targeting the other gentlemen devices

| PD | UD | K = 5 | | K = 10 | | K = 15 | |
|---|---|---|---|---|---|---|---|
| | | Probability | Throughput | Probability | Throughput | Probability | Throughput |
| A | BCD | 0.655 | 0.733353 | 0.712 | 0.713375 | 0.729 | 0.726894 |
| AB | CD | 0.624 | 0.654857 | 0.645 | 0.636341 | 0.829 | 0.646891 |
| ABC | D | 0.887 | 0.684555 | 0.963 | 0.580316 | 0.996 | 0.513519 |
| ACD | B | 0.956 | 0.547798 | 0.986 | 0.346246 | 0.997 | 0.241135 |
| ADB | C | 0.946 | 0.423408 | 0.951 | 0.309469 | 0.972 | 0.267136 |
| AC | BD | 0.964 | 0.772681 | 0.979 | 0.662404 | 0.988 | 0.582111 |
| AD | CB | 0.461 | 0.739893 | 0.415 | 0.690475 | 0.381 | 0.687251 |
| B | ACD | 0.496 | 0.741337 | 0.487 | 0.723239 | 0.465 | 0.717923 |
| BC | AD | 0.817 | 0.713618 | 0.899 | 0.617848 | 0.959 | 0.547188 |
| BCD | A | 0.938 | 0.520275 | 0.968 | 0.3279814 | 0.985 | 0.228197 |
| BD | AC | 0.927 | 0.454418 | 0.927 | 0.362362 | 0.954 | 0.313664 |
| C | ABD | 0.901 | 0.843735 | 0.91 | 0.72117 | 0.959 | 0.663217 |
| CD | AB | 0.897 | 0.555124 | 0.919 | 0.38078 | 0.939 | 0.279741 |
| D | ABC | 0.859 | 0.610293 | 0.878 | 0.536433 | 0.862 | 0.535952 |

As can be expected one simple solution to increase both the throughput and the security is to increase device battery or upgrade processing power. An alternative solution may be to replace the *critical* devices with more potent ones. However assuming the device specifications are constant and the only customization is the strength of the puzzles and which devices to protect, through our tool we are able to identify what produces the best results. The topic of battery adds a further layer to the common question of security vs output and creates several additional layers of complexity. What we have gathered from our analysis and results is in line with our initial hypothesis. Furthermore, this model enables the discovery of *critical* sections in an IoT system which might not be as easy to find compared to our simplified case study.

## 7 Conclusion

We have presented a model that provides means to quantify the effectiveness of a DoS threat on a potential IoT system. Our methodology enables for a potential user to make informed decisions regarding the potential implementation of their system. By using a combination of verification and simulation we confirmed the hypothesis that for the usage in IoT some mitigation techniques could potentially cause harm, however thanks to this tool, one can now make sure to maximize their systems effectiveness. Our future work will look into applying this model to a real world implementation to evaluate the effectiveness of the analysis. We also intend to evaluate different mitigation techniques and specific IoT protocols. Further experimentation could also assess different kinds of IoT devices. We have used the term IoT node to represent a low powered device. However further tests could also examine the best kind of device for balancing their processing power against effectiveness.